\documentclass[preprint,prd]{revtex4}
\usepackage{amssymb}
\usepackage{amsmath}
\usepackage{graphicx}

\input{tcilatex}
\begin{document}

\title[]{Interaction of a Magnet and a Point Charge: Unrecognized Internal
Electromagnetic Momentum Eliminates the Myth of Hidden Mechanical Momentum}
\author{Timothy H. Boyer}
\affiliation{Department of Physics, City College of the City University of New York, New
York, New York 10031}

\begin{abstract}
A model calculation using the Darwin Lagrangian is carried out for a magnet
consisting of two current-carrying charges constrained by centripetal forces
to move in a circular path in the presence of the electric field from a
distant external point charge. \ In the limit that the magnet's two charges
are \textit{non-interacting}, the calculation recovers the only valid
calculation for \textquotedblleft hidden \textit{mechanical}
momentum.\textquotedblright\ \ However, if the magnet's charges are mutually 
\textit{interacting}, then there is internal \textit{electromagnetic} linear
momentum associated with the perturbed magnet's electrostatic charge
distribution and the motion of the magnet's charges. \ This internal \textit{%
electromagnetic} momentum does not seem to be recognized as distinct from
the familiar external electromagnetic momentum associated with the electric
field of the external charge and the magnetic field of the unperturbed
magnet. \ In the multiparticle limit, the \textquotedblleft hidden \textit{%
mechanical} momentum\textquotedblright\ becomes negligible while the
internal \textit{electromagnetic} momentum provides the compensating linear
momentum required by the relativistic conservation law connecting the total
linear momentum to motion of the center of energy. \ Whereas the changes in
the external electromagnetic momentum are often associated with \textit{%
magnetic} forces of order $1/c^{2},$ changes in the internal electromagnetic
momentum are associated with the \textit{electrical} forces of order $%
1/c^{2}.$ \ These\ electrical forces are relevant to the Shockley-James
paradox and to the experimentally observed Aharonov-Bohm and Aharonov-Casher
phase shifts.
\end{abstract}

\maketitle

\section{Introduction}

\subsection{Problem of the Interaction of a Magnet and a Point Charge}

Ever since Aharonov and Bohm's claim\cite{AB} that a charged particle could
be influenced by magnetic fluxes in the absence of electromagnetic forces,
there has been increased interest in the interaction of a magnet and a point
charge. \ In connection with this interaction, Shockley and James\cite{SJ}
introduced the idea of \textquotedblleft hidden momentum\textquotedblright\
and Coleman and van Vleck\cite{CV} suggested this \textquotedblleft hidden
momentum\textquotedblright\ was entirely mechanical in nature. \ The idea of
\textquotedblleft hidden mechanical momentum\textquotedblright\ now appears
in leading textbooks of electromagnetism.\cite{G}\cite{J}

However, despite an extensive literature invoking \textquotedblleft hidden
mechanical momentum,\textquotedblright \cite{resource} there still remains
only one valid calculation for this quantity in the literature. \ This
calculation involves a current loop modeled as a single charged particle (or
equivalently many \textit{non-interacting} particles) moving in a fixed
orbit under the influence of an external electric field generated by the
external point charge. \ This one-particle calculation was given by Penfield
and Haus,\cite{PH} is sketched in a foot note in the work of Coleman and van
Vleck,\cite{CV} and today appears as an example\cite{G520} in an excellent
undergraduate electromagnetism textbook.\ All the remaining references to
hidden momentum represent empty claims unsupported by any calculation.

In the present article, we move beyond the one-moving-particle magnet over
to a two-\textit{interacting}-moving-particle magnet. \ Just as the
one-particle magnet allows a complete calculation of the system linear
momentum, so too this two-interacting-moving-particle magnet allows complete
calculation of the system linear momentum using the Darwin Lagrangian valid
to order $v^{2}/c^{2}$. \ The \textit{interacting} multiparticle model for
the magnet introduces an aspect which does not seem to have been recognized
previously in the literature. \ The mutual interactions of the charges of
the magnet in the perturbing presence of the external electric field produce
an internal \textit{electromagnetic} linear momentum. \ In the limit of a
large number of charges (where the electromagnetic interactions dominate the
mechanical aspects of the magnet's behavior), the hidden \textit{mechanical}
momentum of the noninteracting-particle magnet vanishes and is replaced by
an equal amount of internal \textit{electromagnetic} momentum. \ Thus the
physicists who insisted upon a vanishing total linear momentum for a
relativistic system with a stationary center of energy (and vanishing power
from external forces) are indeed noting a demand of relativistic theory;
however, the identification of the unrecognized linear momentum as "\textit{%
mechanical}" seems entirely misdirected. \ Just as the \textit{mechanical}
kinetic energy in a current-carrying solenoid can essentially always be
ignored compared to the \textit{electromagnetic} energy, so the \textit{%
mechanical} linear momentum of a multiparticle magnetic can essentially
always be ignored compared to the internal \textit{electromagnetic} linear
momentum in the presence of an external electric field.

In an attempt to correct the misdirected hidden-mechanical-momentum claims
in the electromagnetism literature, the present article presents the
elementary calculation for the internal electromagnetic momentum of a
two-interacting-moving-particle magnet in the presence of an external
electric field due to an external point charge. \ The change from hidden 
\textit{mechanical} momentum over to internal \textit{electromagnetic}
momentum has crucial implications for the forces on a charged particle
passing a solenoid and hence upon the proper interpretation of the
experimental observations for this interaction. \ However, the discussion of
the unrecognized classical electromagnetic forces responsible for the
Aharonov-Bohm\cite{AB} and Aharonov-Casher\cite{AC} phase shifts will be
reserved for another publication.

\subsection{Outline of the Presentation}

We start our analysis by reviewing the Darwin Lagrangian and the expression
for the canonical momentum associated with each charged particle. \ Then we
note the relativistic conservation law connecting the system linear momentum
and the velocity of the system center of energy. \ Having reviewed these
preliminaries, we consider a model for a magnet which has charged particles
constrained by centripetal forces to move in a circular path (corresponding
to charged beads sliding on a frictionless ring) while a neutralizing
particle is held at the center of the circular path. \ A distant external
particle is held at a fixed location by external forces of constraint. \
Next we calculated the electromagnetic field momentum involving the electric
field of the external charge and the magnetic field of the unperturbed
moving charges of the magnet. \ After this calculation for the unperturbed
situation, we consider the perturbed motion of the magnet charges due to the
presence of the external charge. \ First we review the case of a
one-moving-particle magnet (or equivalently many \textit{non-interacting}
magnet charges). \ We repeat the earlier calculation in the literature
showing that the mechanical momentum of the magnet particle is equal in
magnitude and opposite in direction compared to the electromagnetic field
momentum between the charge and the unperturbed magnet. \ Thus we see that
the relativistic conservation law regarding the center of energy indeed
holds. \ Then we turn to the model for a magnet which does not seem to have
been treated earlier. \ We consider two \textit{interacting} charges and
calculate the steady-state perturbed motion in the presence of the external
charge. \ We find that the internal linear moment of the magnet in the
presence of the external charge now has two contributions, a mechanical part
and an electromagnetic part; the electromagnetic part is due to the motion
of each of the charged magnet particles in the field of the other magnet
particle. \ The relativistic conservation law is again observed. \ However,
it is clear that as the size of the electrical charges increases, the
electromagnetic momentum contribution increases and the mechanical
contribution decreases. \ Indeed in the case of a large number of
interacting magnet particles, the internal electromagnetic linear momentum
in the presence of the external charge will overwhelm the mechanical
momentum. \ In a multiparticle magnet, the \textquotedblleft hidden
mechanical momentum\textquotedblright\ is completely negligible. \ Finally
we discuss the idea of internal electromagnetic momentum in connection with
the claims in the textbook and research literature. \ We point out that
internal electromagnetic momentum will lead to forces on the external charge
which are completely different from those associated with \textquotedblleft
hidden mechanical momentum.\textquotedblright\ \ The recognition of these
classical electromagnetic forces is important for understanding the
experimentally observed Aharonov-Bohm and Aharonov-Casher phase shifts and
also the role of resistivity in interactions between current loops and
passing charges.

\section{Basis for the Calculations}

\subsection{The Darwin Lagrangian}

Although the experimentally-measured Aharonov-Bohm phase shift involving the
interaction of a magnet and a point charge is usually described within 
\textit{non-relativistic} quantum theory,\cite{Gq} the effect is actually of
order $1/c^{2}$. \ In SI units the phase shift is $q\Phi /\hbar ,$ whereas
in gaussian units the phase shift is $q\Phi /(\hbar c)$ with the magnetic
flux $\Phi $ contributing a second factor of $1/c.$ \ Thus we expect that
the interaction of a magnet and a point charge needs to be treated
relativistically, at least to order $v^{2}/c^{2}$. \ Within classical
electrodynamics, the electromagnetic interaction of point charges $e_{i}$ at
locations $\mathbf{r}_{i}$ moving with velocity $\mathbf{v}_{i}$ can be
described through order $v^{2}/c^{2}$ by the Darwin Lagrangian\cite{DL}

\begin{eqnarray}
\mathcal{L} &\mathcal{=}&\tsum\limits_{i=1}^{i=N}m_{i}c^{2}\left( -1+\frac{%
\mathbf{v}_{i}^{2}}{2c^{2}}+\frac{(\mathbf{v}_{i}^{2})^{2}}{8c^{4}}\right) -%
\frac{1}{2}\tsum\limits_{i=1}^{i=N}\tsum\limits_{j\neq i}\frac{e_{i}e_{j}}{|%
\mathbf{r}_{i}-\mathbf{r}_{j}|}  \notag \\
&&+\frac{1}{2}\tsum\limits_{i=1}^{i=N}\tsum\limits_{j\neq i}\frac{e_{i}e_{j}%
}{2c^{2}}\left[ \frac{\mathbf{v}_{i}\cdot \mathbf{v}_{j}}{|\mathbf{r}_{i}-%
\mathbf{r}_{j}|}+\frac{\mathbf{v}_{i}\cdot (\mathbf{r}_{i}-\mathbf{r}_{j})%
\mathbf{v}_{j}\cdot (\mathbf{r}_{i}-\mathbf{r}_{j})}{|\mathbf{r}_{i}-\mathbf{%
r}_{j}|^{3}}\right]  \label{e1}
\end{eqnarray}

\subsection{Electric and Magnetic Fields from the Darwin Lagrangian}

The Lagrangian equations of motion following from Eq. (\ref{e1}) can be
rewritten in the form of Newton's second law for the mechanical momentum $%
\mathbf{p}^{mech}$ and force $\mathbf{F,}$ $d\mathbf{p}^{mech}/dt=d(m\gamma 
\mathbf{v})/dt=\mathbf{F,}$ with $\gamma =(1-v^{2}/c^{2})^{-1/2}.$ \ In this
Newtonian form, we have%
\begin{eqnarray}
&&\frac{d}{dt}\left[ m_{i}\gamma _{i}\mathbf{v}_{i}\right] =\frac{d}{dt}%
\left[ \frac{m_{i}\mathbf{v}_{i}}{(1-\mathbf{v}_{i}^{2}/c^{2})^{1/2}}\right]
\approx \frac{d}{dt}\left[ m_{i}\left( 1+\frac{\mathbf{v}_{i}^{2}}{2c^{2}}%
\right) \mathbf{v}_{i}\right]  \notag \\
&=&e_{i}\tsum\limits_{j\neq i}\mathbf{E}_{j}(\mathbf{r}_{i},t)+e_{i}\frac{%
\mathbf{v}_{i}}{c}\times \tsum\limits_{j\neq i}\mathbf{B}_{j}(\mathbf{r}%
_{i},t)  \label{e5}
\end{eqnarray}%
with the Lorentz force on the $i$th particle arising from the
electromagnetic fields of the other particles. \ The electromagnetic fields
due to the $j$th particle are given through order $v^{2}/c^{2}$ by\cite{PA}%
\begin{eqnarray}
\mathbf{E}_{j}(\mathbf{r,t}) &=&e_{j}\frac{(\mathbf{r}-\mathbf{r}_{j})}{|%
\mathbf{r}-\mathbf{r}_{j}|^{3}}\left[ 1+\frac{\mathbf{v}_{j}^{2}}{2c^{2}}-%
\frac{3}{2}\left( \frac{\mathbf{v}_{j}\cdot (\mathbf{r}-\mathbf{r}_{j})}{c|%
\mathbf{r}-\mathbf{r}_{j}|}\right) ^{2}\right]  \notag \\
&&-\frac{e_{j}}{2c^{2}}\left( \frac{\mathbf{a}_{j}}{|\mathbf{r}-\mathbf{r}%
_{j}|}+\frac{\mathbf{a}_{j}\cdot (\mathbf{r}-\mathbf{r}_{j})(\mathbf{r}-%
\mathbf{r}_{j})}{|\mathbf{r}-\mathbf{r}_{j}|^{3}}\right)  \label{e6}
\end{eqnarray}%
and 
\begin{equation}
\mathbf{B}_{j}(\mathbf{r},t)=e_{j}\frac{\mathbf{v}_{j}}{c}\times \frac{(%
\mathbf{r}-\mathbf{r}_{j})}{|\mathbf{r}-\mathbf{r}_{j}|^{3}}  \label{e7}
\end{equation}%
where in Eq. (\ref{e6}) the quantity $\mathbf{a}_{j}$ refers to the
acceleration of the $j$th particle.

\subsection{Canonical Linear Momentum from the Darwin Lagrangian}

The canonical linear momentum $\mathbf{p}_{i}=\mathbf{p}_{i}^{mech}+\mathbf{p%
}_{i}^{field}$ associated with the $i$th charge is given by%
\begin{equation}
\frac{\partial \mathcal{L}}{\partial \mathbf{v}_{i}}=\mathbf{p}%
_{i}=m_{i}\left( 1+\frac{\mathbf{v}_{i}^{2}}{2c^{2}}\right) \mathbf{v}%
_{i}+\tsum\limits_{j\neq i}\frac{e_{i}e_{j}}{2c^{2}}\left( \frac{\mathbf{v}%
_{j}}{|\mathbf{r}_{i}-\mathbf{r}_{j}|}+\frac{\mathbf{v}_{j}\cdot (\mathbf{r}%
_{i}-\mathbf{r}_{j})(\mathbf{r}_{i}-\mathbf{r}_{j})}{|\mathbf{r}_{i}-\mathbf{%
r}_{j}|^{3}}\right)  \label{e2}
\end{equation}%
corresponding to the \textit{mechanical} linear momentum of the $i$th
particle%
\begin{equation}
\mathbf{p}_{i}^{mech}=m_{i}\gamma _{i}\mathbf{v}_{i}\approx
m_{i}[1+v_{i}^{2}/(2c^{2})]\mathbf{v}_{i}=m_{i}\mathbf{v}_{i}+m_{i}v_{i}^{2}%
\mathbf{v}_{i}/(2c^{2})  \label{e2a}
\end{equation}%
and the \textit{electromagnetic} linear momenta associated with the \textit{%
electric} field of the $i$th particle and \textit{magnetic} fields of the
other $j$th particles%
\begin{equation}
\mathbf{p}_{i}^{field}=\dsum\limits_{j\neq i}1/(4\pi c)\tint d^{3}r\mathbf{E}%
_{i}\times \mathbf{B}_{j}=\tsum\limits_{j\neq i}\frac{e_{i}e_{j}}{2c^{2}}%
\left( \frac{\mathbf{v}_{j}}{|\mathbf{r}_{i}-\mathbf{r}_{j}|}+\frac{\mathbf{v%
}_{j}\cdot (\mathbf{r}_{i}-\mathbf{r}_{j})(\mathbf{r}_{i}-\mathbf{r}_{j})}{|%
\mathbf{r}_{i}-\mathbf{r}_{j}|^{3}}\right)  \label{e2b}
\end{equation}%
\ Thus the total linear momentum $\mathbf{P}$ is given by 
\begin{equation}
\mathbf{P}=\dsum\limits_{i}\mathbf{p}_{i}=\tsum\limits_{i}m_{i}\left( 1+%
\frac{\mathbf{v}_{i}^{2}}{2c^{2}}\right) \mathbf{v}_{i}+\frac{1}{2}%
\tsum\limits_{i}\tsum\limits_{j\neq i}\frac{e_{i}e_{j}}{2c^{2}}\left( \frac{%
\mathbf{v}_{j}}{|\mathbf{r}_{i}-\mathbf{r}_{j}|}+\frac{\mathbf{v}_{j}\cdot (%
\mathbf{r}_{i}-\mathbf{r}_{j})(\mathbf{r}_{i}-\mathbf{r}_{j})}{|\mathbf{r}%
_{i}-\mathbf{r}_{j}|^{3}}\right)  \label{e3}
\end{equation}

\subsection{Relativistic Conservation Laws}

The Darwin Lagrangian satisfies the familiar conservation laws involving
energy, linear momentum, and angular momentum. \ Through order $v^{2}/c^{2}$
the Darwin Lagrangian satisfies the last (and only specifically
relativistic) conservation law for relativistic systems involving the
uniform motion of the center of energy. \ This last relativistic
conservation law takes the general form\cite{B2005}%
\begin{equation}
\tsum\limits_{i}(\mathbf{F}_{ext\text{ }i}\cdot \mathbf{v}_{i})\mathbf{r}%
_{i}=\frac{d}{dt}\left( U\overrightarrow{\mathcal{X}}_{E}\right) -c^{2}%
\mathbf{P}  \label{e8}
\end{equation}%
connecting the external forces $\mathbf{F}_{ext\text{ }i}$ on the particles
to the time-rate-of-change of the system energy $U$ times the center of
energy $\overrightarrow{\mathcal{X}}_{E}$ and the total system momentum $%
\mathbf{P}$. \ We see that when there is no power introduced by external
forces on the system (so that the the left-hand side of Eq. (\ref{e8})
vanishes and the system energy $U$ is conserved), then the relativistic
conservation law reads 
\begin{equation}
Ud\overrightarrow{\mathcal{X}}_{E}/dt=c^{2}\mathbf{P}  \label{e9}
\end{equation}%
In this case, if the center of energy is at rest, $d\overrightarrow{\mathcal{%
X}}_{E}/dt=0,$ then the total system momentum $\mathbf{P}$ must vanish. \ It
is this conservation law to which physicists appeal when introducing the
idea of hidden mechanical momentum. \ Indeed, Jackson has a problem\cite%
{J618} in his graduate text which considers a point charge at the center of
a toroidal magnet. \ There are no external forces on the system, the center
of energy is not moving, and therefore the total linear momentum of the
system must vanish. \ The text then suggests that a hidden mechanical
momentum is present in the magnet. \ Also, in his undergraduate text,
Griffiths has an example\cite{G520} of a rectangular current loop in an
external electric field, and again suggests that hidden mechanical momentum
is present in the current loop. \ Although both these examples involve
correct calculations, the magnets are treated as though they were
one-moving-particle magnets (or \textit{non-interacting}-moving-particle
magnets) since there is no interaction allowed between the particles of the
magnets. \ This crucial restriction to non-interacting particles is contrary
to the whole spirit of multiparticle electromagnetism where the mutual
interactions between charges lead to such behaviors as electrostatic
screening and self-induction.

\section{Basic Model for the Magnet-Point Charge Interaction}

\subsection{Magnet Modeled by Moving Point Charges}

The model for a magnet used in our calculations involves $N$ charges $e_{i}$
of charge $e$ which are held by external centripetal forces of constraint in
a circular orbit of radius $R$ centered on the origin in the $xy$-plane
while an opposite neutralizing charge $Q=$ $-Ne$ is located at the origin. \
The charges $e_{i}$ in the circular orbit are free to move along the orbit
due to any electromagnetic forces which are tangential to the orbit. \ Thus
in essence, the magnetic is pictured as charged beads $e$ sliding on a
frictionless ring in the $xy$-plane with a balancing opposite charge $Q$
located at the center of the ring. \ 

In the absence of any perturbing influence, the charges $e_{i}$ can be
equally spaced with initial phases $\phi _{i}$ around the circular orbit and
move with angular velocity $\omega _{0}$, speed $v_{0}=\omega _{0}R,$
displacement%
\begin{equation}
\mathbf{r}_{0i}=R[\widehat{x}\cos (\omega _{0}t+\phi _{i})+\widehat{y}\sin
(\omega _{0}t+\phi _{i})]=\widehat{r}_{0i}R  \label{n0}
\end{equation}%
and velocity%
\begin{equation}
\mathbf{v}_{0i}=\omega _{0}R[-\widehat{x}\sin (\omega _{0}t+\phi _{i})+%
\widehat{y}\cos (\omega _{0}t+\phi _{i})]=\widehat{\phi }_{0i}\omega _{0}R
\label{n00}
\end{equation}%
Here the radial and tangential unit vectors for the charge $e_{i}$ are 
\begin{equation}
\widehat{r}_{0i}=\widehat{x}\cos (\omega _{0}t+\phi _{i})+\widehat{y}\sin
(\omega _{0}t+\phi _{i})  \label{n00a}
\end{equation}%
and%
\begin{equation}
\widehat{\phi }_{0i}=-\widehat{x}\sin (\omega _{0}t+\phi _{i})+\widehat{y}%
\cos (\omega _{0}t+\phi _{i})  \label{n00b}
\end{equation}%
The magnetic moment $\overrightarrow{\mu }$ of the $N$-moving-particle
magnet is given by the time-average of $e\mathbf{r\times v}/(2c)$
corresponding to 
\begin{equation}
\overrightarrow{\mu }=\left\langle \tsum\limits_{i=1}^{N}e_{i}\mathbf{r}_{i}%
\mathbf{\times v}_{i}/(2c)\right\rangle =\widehat{z}NeR^{2}\omega _{0}/(2c)
\label{e10}
\end{equation}

We now introduce an external charge $q$ located on the $x$-axis at
coordinate $x_{q}$, $\mathbf{r}_{q}=\widehat{x}x_{q},$ which is held in
place by external forces of constraint$.$ \ If the charge $q$ is far away
from the magnet, $R<<x_{q},$ then the electric field $\mathbf{E}_{q}(\mathbf{%
r})$ near the position of the magnet, $r\approx R<<x_{q},$ is given by 
\begin{equation}
\mathbf{E}_{q}(\mathbf{r})=\frac{q(\mathbf{r-r}_{q})}{\left\vert \mathbf{r-r}%
_{q}\right\vert ^{3}}\approx \frac{q(\mathbf{r-}\widehat{x}x_{q})}{x_{q}^{3}}%
\left( 1+\frac{3\widehat{x}\cdot \mathbf{r}}{x_{q}}\right) =\frac{-\widehat{x%
}q}{x_{q}^{2}}+\frac{q[\mathbf{r-}3\widehat{x}(\widehat{x}\cdot \mathbf{r)]}%
}{x_{q}^{3}}  \label{e11}
\end{equation}%
to order $r/x_{q}.$ \ In this article, we will need only the leading term in
the electric field, $\mathbf{E}_{q}(0)\approx -\widehat{x}q/x_{q}^{2}.$

\subsection{Familiar Electromagnetic Field Momentum for an Unperturbed
Current Loop in an External Electric Field}

In the approximation that the charges $e_{i}$ of the magnet are not
perturbed by the presence of the point charge $q$, we can obtain the linear
momentum in the electromagnetic field as the contribution of the electric
field of the point charge $q$ and the magnetic field arising from the moving
magnet charges $e_{i}.$ \ This correspond to the canonical momentum of the
stationary charge $q$ (which is entirely electromagnetic field momentum),%
\begin{eqnarray}
\mathbf{p}_{q} &=&\mathbf{p}_{q}^{field}=\tsum\limits_{i=1}^{N}\frac{1}{4\pi
c}\tint d^{3}r\mathbf{E}_{q}\times \mathbf{B}_{i}=\tsum\limits_{j=1}^{N}%
\frac{qe_{j}}{2c^{2}}\left( \frac{\mathbf{v}_{j}}{|\mathbf{r}_{q}-\mathbf{r}%
_{j}|}+\frac{\mathbf{v}_{j}\cdot (\mathbf{r}_{q}-\mathbf{r}_{j})(\mathbf{r}%
_{q}-\mathbf{r}_{j})}{|\mathbf{r}_{q}-\mathbf{r}_{j}|^{3}}\right)  \notag \\
&\approx &\tsum\limits_{j=1}^{N}\frac{qe_{j}}{2c^{2}}\left[ \frac{\mathbf{v}%
_{j}}{x_{q}}\left( 1+\frac{\widehat{x}\cdot \mathbf{r}_{j}}{x_{q}}\right) +%
\frac{\mathbf{v}_{j}\cdot (\widehat{x}x_{q}-\mathbf{r}_{j})(\widehat{x}x_{q}-%
\mathbf{r}_{j})}{x_{q}^{3}}\left( 1+\frac{3\widehat{x}\cdot \mathbf{r}_{j}}{%
x_{q}}\right) \right]  \label{e12}
\end{eqnarray}%
\ We now insert the expressions (\ref{n0}) and (\ref{n00}) into Eq. (\ref%
{e12}), average over time, and keep terms through first order $1/x_{q}^{2}.$
\ We note that $\mathbf{v}_{j}\cdot \mathbf{r}_{j}=0,$ $\ \left\langle 
\mathbf{v}_{j}\right\rangle =0,$ $\ \left\langle (\mathbf{v}_{j}\cdot 
\widehat{x})(\widehat{x}\cdot \mathbf{r}_{j})\right\rangle =0,$ \ and \ $%
\left\langle \mathbf{v}_{j}(\widehat{x}\cdot \mathbf{r}_{j})\right\rangle =%
\widehat{y}\omega _{0}R^{2}/2=-\left\langle (\mathbf{v}_{j}\cdot \widehat{x})%
\mathbf{r}_{j}\right\rangle .$ \ The time-averaged electromagnetic field
momentum $\mathbf{p}_{q}$ receives equal contributions from each charge
giving 
\begin{equation}
\left\langle \mathbf{p}_{q}\right\rangle =\frac{\widehat{y}Nqe\omega
_{0}R^{2}}{2c^{2}x_{q}^{2}}=\frac{1}{c}\mathbf{E}_{q}(0)\times 
\overrightarrow{\mu }  \label{e15}
\end{equation}%
This is the familiar expression for the electromagnetic field momentum for
an unperturbed localized steady current located in an external electric
field; it corresponds to a problem in Jackson's text.\cite{J286} \ Even if
the magnet is perturbed by the electric field $\mathbf{E}_{q}(0)$, the field
momentum in Eq. (\ref{e15}) will remain unchanged to lowest order in this
perturbing field.

\section{\textquotedblleft Hidden Mechanical Momentum\textquotedblright\ in
the One-Moving-Particle Magnet}

\subsection{Perturbation of the One-Moving-Particle Magnet}

In order to connect our new work on internal electromagnetic momentum with
the previous literature involving \textquotedblleft hidden mechanical
momentum,\textquotedblright\ we start with a magnet consisting of one moving
charged particle of mass $m$ \ and charge $e;$ for this one-moving-particle
magnet the index \textquotedblleft $i$\textquotedblright\ corresponds only
to $i=1$. \ The neutralizing charge $Q=-e$ is located at the origin. \ (This
one-moving-particle magnet is actually equivalent to assuming that there are 
$N$ \textit{non-interacting} moving particles.) \ 

The position of the charged particle $e$ in the magnet is perturbed by the
presence of the external electric field due to the charge $q.$ The phase
angle is no longer the $\omega _{0}t+\phi _{i}$ appearing in Eq. (\ref{n0}),
but rather becomes \ $\omega _{0}t+\phi _{i}+\eta _{i}(t),$ so that the
displacement of the particle through first order in the perturbation is now 
\begin{equation}
\mathbf{r}_{i}=R\{\widehat{x}\cos [\omega _{0}t+\phi _{i}+\eta _{i}(t)]+%
\widehat{y}\sin [\omega _{0}t+\phi _{i}+\eta _{i}(t)]\}=R\widehat{r}_{i}=%
\mathbf{r}_{0i}+\delta \mathbf{r}_{i}  \label{n2}
\end{equation}%
where%
\begin{equation}
\widehat{r}_{i}=\widehat{x}\cos [\omega _{0}t+\phi _{i}+\eta _{i}(t)]+%
\widehat{y}\sin [\omega _{0}t+\phi _{i}+\eta _{i}(t)]  \label{n2a}
\end{equation}%
the unperturbed displacement $\mathbf{r}_{0i}$ is given in Eq. (\ref{n0}),
and 
\begin{equation}
\delta \mathbf{r}_{i}=\eta _{i}R\{-\widehat{x}\sin [\omega _{0}t+\phi _{i}]+%
\widehat{y}\cos [\omega _{0}t+\phi _{i}]\}=\widehat{\phi }_{0i}R\eta _{i}
\label{nn2}
\end{equation}%
Due to the perturbation, the velocity $\mathbf{v}_{i}=d\mathbf{r}_{i}/dt$ is
now%
\begin{eqnarray}
\mathbf{v}_{i} &=&(\omega _{0}+d\eta _{i}/dt)R\{-\widehat{x}\sin [\omega
_{0}t+\phi _{i}+\eta _{i}(t)]+\widehat{y}\cos [\omega _{0}t+\phi _{i}+\eta
_{i}(t)]\}  \notag \\
&=&\widehat{\phi }_{i}(\omega _{0}+d\eta _{i}/dt)R=\mathbf{v}_{0i}+\delta 
\mathbf{v}_{i}  \label{n3}
\end{eqnarray}%
where%
\begin{equation}
\widehat{\phi }_{i}=-\widehat{x}\sin [\omega _{0}t+\phi _{i}+\eta _{i}(t)]+%
\widehat{y}\cos [\omega _{0}t+\phi _{i}+\eta _{i}(t)]  \label{n3a}
\end{equation}%
the unperturbed velocity $\mathbf{v}_{0i}$ is given in Eq. (\ref{n00}), and 
\begin{equation}
\delta \mathbf{v}_{i}=\widehat{\phi }_{0i}Rd\eta _{i}/dt-\widehat{r}%
_{0i}R\omega _{0}\eta _{i}  \label{nn3}
\end{equation}%
through first order in the perturbation $\eta _{i}.$ \ In obtaining Eqs. (%
\ref{nn2}) and (\ref{nn3}), we have used Eqs. (\ref{n00a}) and (\ref{n00b})
as well as the small- angle approximations $\cos (\phi +\delta \phi )\approx
\cos \phi -\delta \phi \sin \phi $ and $\sin (\phi +\delta \phi )\approx
\sin \phi +\delta \phi \cos \phi .$ \ 

The mechanical momentum in Eq. (\ref{e2a}) involves two terms $\mathbf{p}%
_{i}^{mech}=m_{i}\mathbf{v}_{i}+m_{i}v_{i}^{2}\mathbf{v}_{i}/(2c^{2})$. \
When averaged in time, the first term vanishes, $\left\langle m_{i}\mathbf{v}%
_{i}\right\rangle =m_{i}\left\langle \mathbf{v}_{i}\right\rangle =0,$ since
for a stationary situation the time-average velocity vanishes. \
Consequently, the average mechanical momentum in a stationary situation
involves terms which already contain a factor of $1/c^{2},$ $\left\langle 
\mathbf{p}_{i}^{mech}\right\rangle =\left\langle m_{i}v_{i}^{2}\mathbf{v}%
_{i}/(2c^{2})\right\rangle ,~$ so that the velocity $\mathbf{v}_{i}$ needs
to be calculated only through nonrelativistic order.

\subsection{Calculation of the Perturbation by Energy Conservation}

For the one-moving-particle magnet (or $N$-non-interacting-moving-particle
magnet), it is convenient to obtain the perturbed phase $\eta _{i}(t)$ from
energy conservation. \ The centripetal forces of constraint do no work, and
hence the total energy (kinetic plus electrostatic particle energy) of the
particle (or of each particle of a non-interacting group) is conserved, 
\begin{equation}
\frac{1}{2}mv_{0i}^{2}=\frac{1}{2}mv_{i}^{2}+(eq/x_{q}^{2})R\cos [\omega
_{0}t+\phi _{i}+\eta _{i}(t)]  \label{n4}
\end{equation}%
where the last term is the potential energy of the magnet charge $e$ in the
electrostatic field of the external charge $q$. \ Now we are interested in
the behavior of the system through first order in the perturbation $%
eq/x_{q}^{2}.$ \ Thus we expand each of the terms on the right-hand side of
Eq. (\ref{n4}). \ Expanding the particle kinetic energy $mv^{2}/2,$ we have
from Eqs. (\ref{n00}), (\ref{n3}) and (\ref{nn3})%
\begin{equation}
\frac{1}{2}mv_{i}^{2}=\frac{1}{2}m(\mathbf{v}_{0i}+\delta \mathbf{v)}^{2}=%
\frac{1}{2}mv_{0i}^{2}+m\mathbf{v}_{0i}\cdot \delta \mathbf{v}_{i}=\frac{1}{2%
}mv_{0i}^{2}+m\omega _{0}R^{2}\frac{d\eta _{i}}{dt}  \label{n5a}
\end{equation}%
Also, the term involving the cosine in Eq. (\ref{n4}) is already first order
in the perturbation, and therefore we may drop the $\eta _{i}$ in the
argument of the cosine. \ The energy conservation equation then becomes%
\begin{equation}
\frac{1}{2}mv_{0i}^{2}=\frac{1}{2}mv_{0i}^{2}+m\omega _{0}R^{2}\frac{d\eta
_{i}}{dt}+\frac{eq}{x_{q}^{2}}R\cos (\omega _{0}t+\phi _{i})  \label{n6}
\end{equation}%
or%
\begin{equation}
\frac{d\eta _{i}}{dt}=-\frac{eq}{x_{q}^{2}m\omega _{0}R}\cos (\omega
_{0}t+\phi _{i})  \label{n7}
\end{equation}%
Integrating once, we have the perturbing phase as%
\begin{equation}
\eta _{i}(t)=-\frac{eq}{x_{q}^{2}m\omega _{0}^{2}R}\sin (\omega _{0}t+\phi
_{i})  \label{n8}
\end{equation}

\subsection{Mechanical Momentum of the Perturbed One-Moving-Particle Magnet}

Since the external charge $q$ is not moving, the canonical momentum $\mathbf{%
p}_{e}$ given in Eq. (\ref{e2}) for the one-moving-particle magnet consists
entirely of mechanical momentum. \ We can now calculate the average
mechanical linear momentum of the charge $e$ which is moving in a circular
orbit. \ Through first order in the perturbing force $eq/x_{q}^{2}$ and
second order in $v_{0}/c,$ the mechanical momentum $\mathbf{p}_{m}^{mech}$
is given from Eqs. (\ref{e2a}) and (\ref{n3}) by 
\begin{align}
\mathbf{p}_{e}& =\mathbf{p}_{m}^{mech}=m\gamma _{e}\mathbf{v}_{e}=(m+m%
\mathbf{v}_{0}\cdot \delta \mathbf{v}_{i}/c^{2})(\mathbf{v}_{0i}+\delta 
\mathbf{v}_{i})  \notag \\
& =m(\mathbf{v}_{0i}+\delta \mathbf{v}_{i})+m(\mathbf{v}_{0}\cdot \delta 
\mathbf{v}_{i})\mathbf{v}_{0i}/c^{2}  \label{n9}
\end{align}%
Then averaging in time and noting that $\left\langle \mathbf{v}%
_{0i}\right\rangle =\left\langle \delta \mathbf{v}_{i}\right\rangle =0,$we
have from Eq. (\ref{n00}), (\ref{n8}) and (\ref{n9}), 
\begin{eqnarray}
\mathbf{p}_{e} &=&\left\langle \frac{m\gamma _{0}^{3}(\mathbf{v}_{0}\cdot
\delta \mathbf{v}_{i})\mathbf{v}_{0i}}{c^{2}}\right\rangle =\left\langle 
\frac{m\omega _{0}R^{2}}{c^{2}}\frac{d\eta _{i}}{dt}\mathbf{v}%
_{0i}\right\rangle  \notag \\
&=&\left\langle \frac{m\omega _{0}R^{2}}{c^{2}}\left( -\frac{eq}{%
x_{q}^{2}m\omega _{0}R}\cos (\omega _{0}t+\phi _{i})\right) \omega _{0}R[-%
\widehat{x}\sin (\omega _{0}t+\phi _{i})+\widehat{y}\cos (\omega _{0}t+\phi
_{i})]\right\rangle  \notag \\
&=&-\widehat{y}\frac{eq\omega _{0}R^{2}}{2c^{2}x_{q}^{2}}  \label{n10}
\end{eqnarray}%
This result is just the negative of Eq. (\ref{e15}) when $N=1.$ \ Thus the
average mechanical momentum (for the single charge $e$ of the
one-moving-particle magnet in the presence of the charge $q$) is equal in
magnitude and opposite in sign from the canonical momentum $\mathbf{p}_{q}$
corresponding to the familiar electromagnetic field momentum associated with
the electric field of the external charge $q$ and the magnetic field of the
magnet. \ Thus the total momentum of the system consisting of the charge $q$
and the one-moving-particle magnet indeed vanishes, 
\begin{equation}
\left\langle \mathbf{P}\right\rangle =\left\langle \mathbf{p}%
_{q}\right\rangle +\left\langle \mathbf{p}_{e}\right\rangle =\left\langle 
\mathbf{p}_{q}^{field}\right\rangle +\left\langle \mathbf{p}%
_{e}^{mech}\right\rangle =\widehat{y}\frac{eq\omega _{0}R^{2}}{%
2x_{q}^{2}c^{2}}-\widehat{y}\frac{eq\omega _{0}R^{2}}{2x_{q}^{2}c^{2}}=0
\label{n11}
\end{equation}%
as required by the relativistic conservation law (\ref{e9}). \ The
mechanical linear momentum $\mathbf{p}_{e}$ given in Eq. (\ref{n10}) is what
is identified in the literature as the \textquotedblleft hidden mechanical
momentum.\textquotedblright

\section{Internal Electromagnetic Momentum in the
Two-Interacting--Moving-Particle Magnet}

\subsection{Improved Model of Two Moving Magnet Charges which Interact}

The physical magnets found in nature do not consist of a single charged
particle (or of non-interacting particles) sliding on a frictionless ring. \
Hence we turn to a model consisting of two \textit{interacting} moving
charged particles as an improvement over our one-moving-particle model for a
magnet. \ Now we have two charges $e$ held by external centripetal forces of
constraint in a circular orbit of radius $R$ centered on the origin in the $%
xy$-plane while a neutralizing charge $Q=-2e$ is located at the origin. \
The calculation for the magnetic moment $\overrightarrow{\mu }$ and the
field momentum associated with the canonical momentum $\mathbf{p}_{q}$
follow as in the calculations above with the results in Eqs. (\ref{e10}) and
(\ref{e15}) corresponding to $N=2$. \ 

\subsection{Internal Electromagnetic Field Momentum}

The total linear momentum $\mathbf{P}\ $of our example involves not only the
electromagnetic field momentum $\mathbf{p}_{q}=\mathbf{p}_{q}^{field}$
associated with the canonical momentum of the stationary particle $q$ but
also the canonical momenta $\mathbf{p}_{e1}$ and $\mathbf{p}_{e2}$
associated with the particles of the magnet. \ Since the charge $q$ is at
rest, it has no magnetic field, and hence it does not contribute to the
canonical momentum of $\mathbf{p}_{e1}$ or $\mathbf{p}_{e2}.$ \ However, the
canonical momentum $\mathbf{p}_{e1}$ of the first magnet particle includes
both its mechanical momentum and also the electromagnetic momentum
associated with its own electric field and the magnetic field of the other
moving particle in the magnet. \ Now the \textit{unperturbed} motion of the
magnet charges given in Eqs. (\ref{n0}) and (\ref{n00}) involves no average
linear momentum because the two charges are always moving with opposite
velocities on opposite sides of the circular orbit. \ However, the \textit{%
perturbed} motion will indeed involve net linear momentum for the magnet
particles. \ As soon as the particles of the magnet have mutual
interactions, then the mechanical kinetic energy changes (which provided the
basis for the internal \textit{mechanical} momentum in the
one-moving-particle magnet) are suppressed as energy goes into electrostatic
energy of the interacting particles. \ The internal \textit{mechanical}
momentum of the magnet decreases because of the electrostatic interactions,
and internal \textit{electromagnetic} momentum appears in the internal
electromagnetic fields. \ We will illustrate this situation explicitly for
our example involving the two-interacting--moving-particle magnet.

\subsection{Calculation of the Perturbation Using Nonrelativistic Forces\ }

In order to obtain the internal linear momentum $\mathbf{p}_{e1}+$ $\mathbf{p%
}_{e2}$ of the magnet in the presence of the external electric field $%
\mathbf{E}_{q}$ due to the charge $q,$ we need to calculate the perturbed
motion of the particles $e_{1}$ and $e_{2}.$ \ The perturbed positions of
the two charges $e_{1}$ and $e_{2}$ of the magnet will be written as in Eq. (%
\ref{n2}), where the unperturbed initial phases differ by $\pi ,$ $\phi
_{1}-\phi _{2}=\pi ,$ and where it is again assumed that $\eta _{i}(t)$ is a
small correction. \ The perturbed velocities of the charges are as given in
Eq. (\ref{n3}), and the accelerations then follow as 
\begin{eqnarray}
\mathbf{a}_{i} &=&(\omega _{0}+d\eta _{i}/dt)^{2}R\{-\widehat{x}\cos [\omega
_{0}t+\phi _{i}+\eta _{i}(t)]-\widehat{y}\sin [\omega _{0}t+\phi _{i}+\eta
_{i}(t)]\}  \notag \\
&&+(d^{2}\eta _{i}/dt^{2})R\{-\widehat{x}\sin [\omega _{0}t+\phi _{i}+\eta
_{i}(t)]+\widehat{y}\cos [\omega _{0}t+\phi _{i}+\eta _{i}(t)]\}  \notag \\
&=&-\widehat{r}_{i}(\omega _{0}+d\eta _{i}/dt)^{2}R+\widehat{\phi }%
_{i}(d^{2}\eta _{i}/dt^{2})R  \label{e16b}
\end{eqnarray}%
where $\widehat{r}_{i}$ and $\widehat{\phi }_{i}$ are given in Eqs. (\ref%
{n2a}) and (\ref{n3a}).

Since the magnet charges $e_{1}$ and $e_{2}$ are constrained to move in a
circular orbit, the perturbation of the charges is determined by the
tangential acceleration. \ For charge $e_{i},$ the equation of motion
requires only the electrostatic forces due to the stationary charge $q$ and
the moving charge $e_{j\neq i}.$ \ There are relativistic fields of order $%
1/c^{2}$ in the tangential direction due to the other magnet charges, but
these $1/c^{2}$-corrections will not contribute to average momentum of the
magnet$.$ \ The radial forces are balanced by the forces of constraint. \
Thus from Eq. (\ref{e16b}), we have for the nonrelativistic equation of
motion%
\begin{eqnarray}
m\mathbf{a}_{i}\cdot \widehat{\phi }_{i} &\approx &mR\frac{d^{2}\eta _{i}}{%
dt^{2}}  \notag \\
&=&\widehat{\phi }_{i}\cdot \lbrack e\mathbf{E}_{q}(\mathbf{r}_{i},t)+e%
\mathbf{E}_{j\neq i}(\mathbf{r}_{i},t)]=\widehat{\phi }_{i}\cdot \left( 
\frac{-\widehat{x}eq}{x_{q}^{2}}+\frac{e^{2}(\mathbf{r}_{i}-\mathbf{r}%
_{j\neq i})}{\left\vert \mathbf{r}_{i}-\mathbf{r}_{j\neq i}\right\vert ^{3}}%
\right)  \notag \\
&=&\left( -\widehat{\phi }_{i}\cdot \widehat{x}\frac{eq}{x_{q}^{2}}-\widehat{%
\phi }_{i}\cdot \mathbf{r}_{j\neq i}\frac{e^{2}}{\left\vert \mathbf{r}_{i}-%
\mathbf{r}_{j\neq i}\right\vert ^{3}}\right)  \label{e16c}
\end{eqnarray}%
since $\widehat{\phi }_{i}\cdot \mathbf{r}_{i}=0,$ where the tangential unit
vector is given in Eq. (\ref{n3a}) while $\mathbf{r}_{j\neq i}=\widehat{x}%
R\cos [\omega _{0}t+\phi _{i}+\pi +\eta _{j\neq i}(t)]+\widehat{y}R\sin
[\omega _{0}t+\phi _{i}+\pi +\eta _{j\neq i}(t)].$ \ Then we have 
\begin{eqnarray}
\widehat{\phi }_{i}\cdot \mathbf{r}_{j\neq i} &=&R\{-\sin (\omega _{0}t+\phi
_{i}+\eta _{i})\cos [\omega _{0}t+\phi _{i}+\pi +\eta _{j\neq i}(t)]  \notag
\\
&&+\cos (\omega _{0}t+\phi _{i}+\eta _{i})\sin [\omega _{0}t+\phi _{i}+\pi
+\eta _{j\neq i}(t)]\}  \notag \\
&=&R\sin (\eta _{j\neq i}-\eta _{i}+\pi )=-R\sin (\eta _{j\neq i}-\eta
_{i})\approx R(\eta _{i}-\eta _{j\neq i})  \label{e16d}
\end{eqnarray}%
where we have used the approximation $\sin \phi \approx \phi $ for small $%
\phi .$ \ The separation $\left\vert \mathbf{r}_{i}-\mathbf{r}_{j\neq
i}\right\vert $ between the charges is second order in the perturbation $%
\eta ,$ and so we may write $\left\vert \mathbf{r}_{i}-\mathbf{r}_{j\neq
i}\right\vert \approx 2R$ in Eq. (\ref{e16c}). \ Then the nonrelativistic
equation of motion (\ref{e16c}) for the charge $e_{i}$ through first order
in the perturbation produced by $eq/x_{q}^{2}$ becomes 
\begin{equation}
mR\frac{d^{2}\eta _{i}}{dt^{2}}=\left( \frac{eq}{x_{q}^{2}}\sin (\omega
_{0}t+\phi _{i})-\frac{e^{2}}{(2R)^{3}}R(\eta _{i}-\eta _{j\neq i})\right)
\label{e16f}
\end{equation}%
We notice that since $\phi _{i}-\phi _{j\neq i}=\pi ,$ this equation (\ref%
{e16f}) is odd under the interchange of the two particles. \ Thus for the
steady-state situation, we must have 
\begin{equation}
\eta _{j\neq i}=-\eta _{i}  \label{e16ff}
\end{equation}%
Then the perturbation in the phase $\eta _{i}$ (in steady state) is given by 
\begin{equation}
mR\frac{d^{2}\eta _{i}}{dt^{2}}=\left( \frac{eq}{x_{q}^{2}}\sin (\omega
_{0}t+\phi _{i})-\frac{e^{2}}{(2R)^{2}}\eta _{i}\right)  \label{e16g}
\end{equation}%
with a steady-state solution%
\begin{equation}
\eta _{i}(t)=\frac{eq}{x_{q}^{2}}\frac{\sin (\omega _{0}t+\phi _{i})}{%
[-m\omega _{0}^{2}R+e^{2}/(2R)^{2}]}  \label{e16h}
\end{equation}

If we take the magnitude $e$ of the charges as small (so that we may neglect
the terms in $e^{2}$ involving interactions between the charges), then
equation (\ref{e16h}) agrees exactly with the one-particle-magnet result in
Eq. (\ref{n8}). \ Thus we recover the non-interacting particle result in the
appropriate small-charge-$e$ limit. \ On the other hand, if the magnitude $e$
of the charges becomes large, then according to Eq. (\ref{e16h}) the
electrostatic interaction contribution $e^{2}/(2R)^{2}$ can dominate the
mechanical contribution $-m\omega _{0}^{2}R.$

\subsection{Internal Momentum in the Magnet}

The internal canonical momentum of the magnet in the presence of the
external charge $q$ is given by the sum over the canonical momenta $\mathbf{p%
}_{i}$ in Eq. (\ref{e2}) where the sum includes only the two charges of the
magnet in our model, each with canonical momentum 
\begin{equation}
\mathbf{p}_{ei}=\mathbf{p}_{ei}^{mech}+\mathbf{p}_{ei}^{field}=m\left( 1+%
\frac{\mathbf{v}_{i}^{2}}{2c^{2}}\right) \mathbf{v}_{i}+\frac{e_{i}e_{j\neq
i}}{2c^{2}}\left( \frac{\mathbf{v}_{j\neq i}}{|\mathbf{r}_{i}-\mathbf{r}%
_{j\neq i}|}+\frac{\mathbf{v}_{j\neq i}\cdot (\mathbf{r}_{i}-\mathbf{r}%
_{j\neq i})(\mathbf{r}_{i}-\mathbf{r}_{j\neq i})}{|\mathbf{r}_{i}-\mathbf{r}%
_{j\neq i}|^{3}}\right)  \label{e25}
\end{equation}%
When averaged in time, we expect equal momentum contributions from each
charge. \ The velocity $\mathbf{v}_{i}$ is given in Eq. (\ref{n3}) where $%
\eta _{i}$ is given in Eq. (\ref{e16h}) and its time derivative is%
\begin{equation}
\frac{d\eta _{i}}{dt}=\omega _{0}\frac{eq}{x_{q}^{2}}\frac{\cos (\omega
_{0}t+\phi _{i})}{[-m\omega _{0}^{2}R+e^{2}/(2R)^{2}]}  \label{e26}
\end{equation}

\subsubsection{Mechanical Linear Momentum of a Perturbed Magnet Charge}

Then the mechanical contribution $\mathbf{p}_{i}^{mech}$ to the linear
momentum is%
\begin{eqnarray}
\mathbf{p}_{ei}^{mech} &=&m\left( 1+\frac{v_{i}^{2}}{2c^{2}}\right) \mathbf{v%
}_{i}=m\mathbf{v}_{i}+m\frac{(\mathbf{v}_{0i}+\delta \mathbf{v}_{i})^{2}}{%
2c^{2}}\mathbf{v}_{i}\approx m\mathbf{v}_{i}+m\frac{\mathbf{v}_{0i}^{2}}{%
2c^{2}}\mathbf{v}_{i}+m\frac{\mathbf{v}_{0i}\cdot \delta \mathbf{v}_{i}}{%
c^{2}}\mathbf{v}_{0i}  \notag \\
&=&m\mathbf{v}_{i}+m\frac{\mathbf{v}_{0i}^{2}}{2c^{2}}\mathbf{v}_{i}+\frac{m%
}{c^{2}}\omega _{0}\frac{d\eta }{dt}R^{2}\mathbf{v}_{0i}  \label{g1}
\end{eqnarray}%
from Eqs. (\ref{n3}) and (\ref{nn3}). \ The average value of the velocity is
zero, $\left\langle \mathbf{v}_{i}\right\rangle =0,$ since the magnet-charge
interaction is assumed stationary. \ The required average for $\mathbf{p}%
_{i}^{mech}$ follows from Eqs. (\ref{n00}), (\ref{e26}), and (\ref{g1}) as 
\begin{equation}
\left\langle \mathbf{p}_{ei}^{mech}\right\rangle =\left\langle \frac{m}{c^{2}%
}\omega _{0}\frac{d\eta }{dt}R^{2}\mathbf{v}_{0i}\right\rangle =\widehat{y}%
\frac{m}{c^{2}}\omega _{0}^{2}R^{2}\frac{eq}{x_{q}^{2}}\frac{\omega _{0}R/2}{%
[-m\omega _{0}^{2}R+e^{2}/(2R)^{2}]}  \label{g2}
\end{equation}

\subsubsection{Electromagnetic Linear Momentum Associated with a Perturbed
Magnet Charge}

The electromagnetic contribution $\mathbf{p}_{i}^{field}$\ corresponds to 
\begin{eqnarray}
\mathbf{p}_{ei}^{field} &=&\frac{e_{i}e_{j\neq i}}{2c^{2}}\left( \frac{%
\mathbf{v}_{j\neq i}}{|\mathbf{r}_{i}-\mathbf{r}_{j\neq i}|}+\frac{\mathbf{v}%
_{j\neq i}\cdot (\mathbf{r}_{i}-\mathbf{r}_{j\neq i})(\mathbf{r}_{i}-\mathbf{%
r}_{j\neq i})}{|\mathbf{r}_{i}-\mathbf{r}_{j\neq i}|^{3}}\right)  \notag \\
&=&\frac{e^{2}}{2c^{2}}\left( \frac{\mathbf{v}_{j\neq i}}{|\mathbf{r}_{i}-%
\mathbf{r}_{j\neq i}|}+\frac{\mathbf{v}_{j\neq i}\cdot \mathbf{r}_{i}(%
\mathbf{r}_{i}-\mathbf{r}_{j\neq i})}{|\mathbf{r}_{i}-\mathbf{r}_{j\neq
i}|^{3}}\right)  \label{e27b}
\end{eqnarray}%
since $\mathbf{v}_{j\neq i}\cdot \mathbf{r}_{j\neq i}=0.$ \ The denominator
will involve a distance $2R$ through first order in the perturbation. \ We
need first 
\begin{equation}
\mathbf{v}_{j\neq i}\cdot \mathbf{r}_{i}=v_{ji}\widehat{\phi }_{j\neq
i}\cdot \mathbf{r}_{i}\approx \omega _{0}R^{2}(-2\eta _{i})  \label{e27c}
\end{equation}%
from Eqs. (\ref{e16d}) and (\ref{e16ff}). \ Then from Eqs. (\ref{n00a}), (%
\ref{e16h}), and (\ref{e27c}), the time-average of $(\mathbf{v}_{j\neq
i}\cdot \mathbf{r}_{i})\mathbf{r}_{i}$ becomes 
\begin{equation}
\left\langle (\mathbf{v}_{j\neq i}\cdot \mathbf{r}_{i})\mathbf{r}%
_{i}\right\rangle =\left\langle \omega _{0}R^{2}(-2\eta _{i})\mathbf{r}%
_{i0}\right\rangle =(-2\omega _{0}R^{2})\frac{eq}{x_{q}^{2}}\frac{1}{%
[-m\omega _{0}^{2}R+e^{2}/(2R)^{2}]}\widehat{y}\frac{R}{2}  \label{e27d}
\end{equation}%
and similarly 
\begin{equation}
\left\langle (\mathbf{v}_{j\neq i}\cdot \mathbf{r}_{i})\mathbf{r}_{j\neq
i}\right\rangle =(2\omega _{0}R^{2})\frac{eq}{x_{q}^{2}}\frac{1}{[-m\omega
_{0}^{2}R+e^{2}/(2R)^{2}]}\widehat{y}\frac{R}{2}  \label{e27e}
\end{equation}%
Then from Eq. (\ref{e27b}) the time-average electromagnetic contribution to $%
\mathbf{p}_{ei}$ is 
\begin{eqnarray}
\left\langle \mathbf{p}_{ei}^{field}\right\rangle &=&\frac{e^{2}}{2c^{2}}%
\left[ (-2\omega _{0}R^{2})\frac{eq}{x_{q}^{2}}\frac{2}{[-m\omega
_{0}^{2}R+e^{2}/(2R)^{2}]}\widehat{y}\frac{R}{2}\left( \frac{1}{(2R)^{3}}%
\right) \right]  \notag \\
&=&-\widehat{y}\left( \frac{e^{2}}{(2R)^{2}}\right) \frac{\omega _{0}R^{2}}{%
2c^{2}}\frac{eq/x_{q}^{2}}{[-m\omega _{0}^{2}R+e^{2}/(2R)^{2}]}  \label{e27h}
\end{eqnarray}%
Adding the mechanical contribution in Eq. (\ref{g2}) and the electromagnetic
contribution in Eq. (\ref{e27h}), we find%
\begin{eqnarray}
\left\langle \mathbf{p}_{ei}\right\rangle &=&\left\langle \mathbf{p}%
_{ei}^{mech}\right\rangle +\left\langle \mathbf{p}_{ei}^{field}\right\rangle
=\widehat{y}\frac{m}{c^{2}}\omega _{0}^{2}R^{2}\frac{eq}{x_{q}^{2}}\frac{%
\omega _{0}R/2}{[-m\omega _{0}^{2}R+e^{2}/(2R)^{2}]}  \notag \\
&&-\widehat{y}\left( \frac{e^{2}}{(2R)^{2}}\right) \frac{\omega _{0}R^{2}}{%
2c^{2}}\frac{eq/x_{q}^{2}}{[-m\omega _{0}^{2}R+e^{2}/(2R)^{2}]}  \notag \\
&=&-\frac{\widehat{y}qe\omega _{0}R^{2}}{2c^{2}x_{q}^{2}}  \label{e27ha}
\end{eqnarray}%
The two equal contributions $\left\langle \mathbf{p}_{ei}\right\rangle $ and 
$\left\langle \mathbf{p}_{ej\neq i}\right\rangle $ from the two particles
give the canonical momentum $\left\langle \mathbf{p}_{magnet}\right\rangle $%
\ of the magnet as 
\begin{equation}
\left\langle \mathbf{p}_{magnet}\right\rangle =\left\langle \mathbf{p}_{ei%
\text{ }}\right\rangle +\left\langle \mathbf{p}_{ej\neq i\text{ }%
}\right\rangle =-2\frac{\widehat{y}qe\omega _{0}R^{2}}{2c^{2}x_{q}^{2}}
\label{e27g}
\end{equation}%
But then the canonical momentum $\left\langle \mathbf{p}_{magnet}\right%
\rangle $\ of the two-particle magnet is equal in magnitude and opposite in
sign compared to the canonical momentum $\left\langle \mathbf{p}%
_{q}\right\rangle $ of the external charge $q$ corresponding to $N=2$ \ in
Eq. (\ref{e15}), which was equal to the familiar electromagnetic field
momentum involving the electric field due to $q$ and the magnetic field of
the magnet. \ We see that the relativistic conservation law (\ref{e9})
regarding the center of energy is indeed satisfied and the total momentum of
the system indeed vanishes. \ 

\subsection{Discussion of Internal Electromagnetic Momentum}

Following equation (\ref{e16h}), we noted that in the limit of small value
for the charge $e$ of the magnet particles, the mechanical momentum
dominated the internal momentum of the magnet. \ This mechanical momentum
corresponds to the \textquotedblleft hidden mechanical
momentum\textquotedblright\ of the textbooks and literature. \ However, in
the opposite limit of large charge $e$ for the magnet particles, the
electromagnetic momentum becomes large and the mechanical momentum becomes
negligible. \ As more particles of fixed mass $m$ and charge $e$ are added
to the magnet while keeping the magnetic moment $\overrightarrow{\mu }$
fixed, the speed $v_{0i}=\omega _{0}R$ of the current carriers becomes ever
smaller so that the mechanical momentum becomes insignificant compared to
the internal electromagnetic momentum. \ Thus for any physical multiparticle
magnet with its enormous number of charge carriers, we expect that only the
internal electromagnetic momentum needs to be considered. \ This internal
electromagnetic momentum is equal in magnitude and opposite in direction
from the electromagnetic field momentum which is found in the elementary
textbook calculations involving a point charge and a steady current. \ The
negligible contribution of the particle mechanical momentum is analogous to
the negligible contribution of the particle kinetic energy to the
self-inductance of a circuit\cite{self} where the mass and the charge of the
charge carriers is never mentioned in the textbooks.

Although Coleman and Van Vleck\cite{CV} insisted that the \textquotedblleft
hidden momentum\textquotedblright of Shockley and James\cite{SJ} was purely
mechanical, other authors (such as Furry\cite{F}) have been more cautious. \
In the literature, \textquotedblleft hidden mechanical
momentum\textquotedblright\ is invoked simply as what is left over after the
electromagnetic momentum has been calculated. However, there seems to be no
recognition of the separation between the external and internal
electromagnetic momentum for a magnet interacting with a point charge. The
external electromagnetic momentum is familiar from all the textbooks as the
momentum associated with the electric field of the external charge and the
magnetic field of the unperturbed magnet. The unrecognized internal
electromagnetic momentum involves the field momentum between the electric
and magnetic fields of the magnet particles undergoing perturbed motion due
to the electric field of the external point charge.

To some readers the shift presented here from a \textquotedblleft hidden
mechanical momentum\textquotedblright\ over to an \textquotedblleft internal
electromagnetic momentum\textquotedblright\ may seem like a distinction
without a difference. \ In both cases, we are describing linear momentum
which is internal to the magnet. \ The crucial difference arises in the
implications of the two different forms of momentum. \ Thus
\textquotedblleft hidden mechanical momentum\textquotedblright\ has been
used as an excuse to claim that a charge passing a magnet experienced no
forces\cite{AR} (which forces might lead to an electromagnetic lag effect),
and has been used to claim that a magnetic moment passing an electric line
charge (which magnetic moment experienced an obvious Lorentz force)
nevertheless moved as though it experienced no force at all.\cite{APV}\cite%
{V} \ These claims have been accepted into the mainstream physics literature
both in research and in the textbooks. \ There seems to be no recognition
that changes in the external and internal electromagnetic momenta are
associated with different types of forces. Changes in the external
electromagnetic momentum are associated with \textit{magnetic} forces of
order $1/c^{2}$. Changes in the internal electromagnetic momentum are
associated with \textit{electrical} forces of order $1/c^{2}$. It is these
electrical forces associated with changes in the internal electromagnetic
momentum which play a crucial role in the experimentally observed
Aharonov-Bohm and Aharonov-Casher phase shifts.\ \ \ \ A discussion of the
classical electromagnetic forces responsible for the Aharonov-Bohm phase
shift and the Aharonov-Casher phase shift will be given in another
publication.\cite{classical} \ Here we note simply that our clarification
involving the myth of \textquotedblleft hidden mechanical
momentum\textquotedblright\ alters our fundamental understanding of one of
the connections between classical and quantum theories.\cite{four}\

\end{document}